# Globalisation of science in kilometres

Ludo Waltman,[1] Robert J.W. Tijssen[1,2] and Nees Jan van Eck[1]

[1] Centre for Science and Technology Studies (CWTS), Leiden University
PO Box 905, 2300 AX Leiden, The Netherlands
{waltmanlr, tijssen, ecknjpvan}@cwts.leidenuniv.nl

[2] Centre for Research on Evaluation, Science and Technology (CREST), Stellenbosch University
Stellenbosch, South Africa

The ongoing globalisation of science has undisputedly a major impact on how and where scientific research is being conducted nowadays. Yet, the big picture remains blurred. It is largely unknown where this process is heading, and at which rate. Which countries are leading or lagging? Many of its key features are difficult if not impossible to capture in measurements and comparative statistics. Our empirical study measures the extent and growth of scientific globalisation in terms of physical distances between co-authoring researchers. Our analysis, drawing on 21 million research publications across all countries and fields of science, reveals that contemporary science has globalised at a fairly steady rate during recent decades. The average collaboration distance per publication has increased from 334 kilometres in 1980 to 1553 in 2009. Despite significant differences in globalisation rates across countries and fields of science, we observe a pervasive process in motion, moving towards a truly interconnected global science system.

## 1. Introduction

Increasingly, our 'planet science' is populated by PhD graduates, scholars, engineers and professors who communicate and collaborate with foreign colleagues on a regular basis (Royal Society, 2011). The physical distance between research partners has become increasingly irrelevant in contemporary science owing to cheap travel, improved ICT facilities and the rise of English as the common language in mainstream science. Part of this pervasive process is driven by the internationalisation ('globalisation') of the higher education markets (Wildavsky, 2010), part by international competition among research universities and academic researchers within the leading science nations (Mohrman, Mab, & Baker, 2008). However, the geographical expansion of academic science is not just about free flows of minds and ideas, or getting linked into dense and interconnected partnership networks. It also relates to the way in which research focuses on particular global issues and problems in the context of changing societal needs and social contracts (Samarasekera, 2009), concentration and agglomeration effects that dominate the economic topography of our world (Florida, 2005), and incentives provided by state-funded initiatives such the European Commission's Framework Programmes (Hoekman, Frenken, & Tijssen, 2010).

Clearly, this diversity of underlying processes and driving forces makes it difficult to gauge or measure, in a comprehensive and systematic way, structural features of scientific globalisation. This analytical set-back is partially solved by an abundance of empirical evidence on the geographical distribution of research activity across the globe (Peters, 2009; Veugelers, 2010). The march of globalisation through the landscape of science is documented by research publications in the open scientific literature, more specifically, by the paper trail left behind in the author addresses of jointly authored publications. The growth rates of publication output over the past decades provide detailed information as to which new physical locations have emerged, and how much these new entrants contribute to the connectivity within worldwide science. Several Asian and South American countries have become prominent contributors, crowned by the spectacular ascendancy of China.

Despite numerous case studies on these topics, virtually nothing is known about the impact of globalisation on the physical distance between research partners and the interconnectedness of science. Is the physical distance between collaborating researchers still increasing, and at what rate? Which fields of science are affected most? And which countries or regions are leading the process or lagging behind? These questions can now be addressed systematically by data-mining the millions of author addresses in co-publications. The methods that we introduce provide unobtrusive distance-based measurements of globalisation processes within and across national borders for all countries and all sciences worldwide.

## 2. Methods

Our empirical analysis can be nested within the research programme of spatial scientometrics (Frenken, Hardeman, & Hoekman, 2009) and builds on a rapidly expanding body of scientometric studies in which internationalisation and globalisation processes in science are examined (Glänzel, 2001; Luukkonen, Tijssen, Persson, & Sivertsen, 1993; Narin, Stevens, & Whitlow, 1991). Most of these studies focus on analysing international co-publications. We take a different approach and study geographical distances between co-authoring researchers.[1] The use of geographical distances is not very common in the literature (Hoekman et al., 2010; Hoekman, Frenken, & Van Oort, 2009; Katz, 1994; Liang & Zhu, 2002; Yan & Sugimoto, in press), and at the scale presented in this paper geographical distances have not been analysed before.

### 2.1. Geocoding procedure

Our analysis is based on publications indexed in the CWTS version of the Web of Science (WoS) database, produced by Thomson Reuters. We selected all WoS publications that were published between 1980 and 2009, that are of the document type 'article' or 'review' and that have at least one author affiliate address. There are 21.4 million publications that satisfy these three criteria. For each of the selected publications, an attempt was made to find the geographical coordinates (i.e., the latitude and the longitude) of the addresses mentioned in the publication's address list.[2] Finding the geographical coordinates of an address is referred to as geocoding.

We employed the following geocoding procedure (cf. Leydesdorff & Persson, 2010). First, all 39.0 million addresses of the selected publications were reduced to a city and a country.[3] Other address elements, such as organisation names, streets and postal codes, were disregarded. Next, for each unique address, the number of times it occurs in the address lists of the selected publications was counted. Performing geocoding for all unique addresses turned out to be infeasible, and we therefore restricted our attention to about 11 000 addresses that occur most frequently. The remaining addresses were not taken into account in the geocoding procedure, and their coordinates were considered unknown. For the selected addresses, coordinates were obtained using the website www.gpsvisualizer.com/geocoder/. This website relies on geocoding services provided by Google and Yahoo. Comparing the two services, we found that they sometimes yield quite different results and also that they sometimes fail to recognise an address. Furthermore, although both services make errors, Google seemed to be somewhat more accurate than Yahoo. Based on these observations, we decided to take the following approach. For each address, the Google-Yahoo distance was calculated, that is, the distance between the coordinates provided by Google and the coordinates provided by Yahoo. An address was verified manually if the Google-Yahoo distance is larger than 50 km and the address occurs more than 200 times in the address lists of the selected publications. In some cases, the verification of an address revealed that both the coordinates of Google and the coordinates of Yahoo were incorrect. Usually, the correct coordinates could then be found manually, but in a small number of cases the correct coordinates remained unknown. An address was also verified manually if the Google-Yahoo distance is larger than 100 km and the address occurs less than 200 times. In these cases, however, the verification of an address was done in a more cursory way. If the correctness of the coordinates of either Google or Yahoo could not be easily established, the coordinates of an address were simply considered unknown. Addresses that did not satisfy one of the above two criteria (about 90% of the selected addresses) were not verified manually. For these addresses, the coordinates provided by Google were taken as the correct ones. In the end, our geocoding procedure yielded coordinates for 98.6% of the 39.0 million addresses of the selected publications.

---

[1] More precisely, we study geographical distances between addresses mentioned in the address lists of publications. When a paper has multiple addresses, this is usually due to co-authorship. However, it is important to keep in mind that authors sometimes have more than one affiliation. This may also cause papers to have multiple addresses (Katz & Martin, 1997). For instance, looking at single-author publications in 2009, it turns out that about 10% has more than one address.

[2] The WoS database distinguishes between the ordinary addresses associated with the authors of a publication and the so-called reprint address of a publication. We disregarded the reprint addresses of all publications that appeared after 1997. Starting from 1998, the reprint address of a publication is usually also mentioned in the ordinary address list of the publication. When the reprint address is not mentioned in the ordinary address list, it seems that in most cases the corresponding author of the publication moved to a new organisation after the research reported in the publication was finished.

To assess the accuracy of our geocoding procedure, we manually verified the coordinates of a limited number of addresses. Out of the 11 000 addresses that were taken into consideration in the geocoding procedure, a random sample of 150 addresses was taken. For each of the 150 addresses, we manually identified the geographical coordinates. We then compared the manually identified coordinates with the coordinates obtained using the geocoding procedure. There turned out to be four addresses for which the distance between the manually identified coordinates and the geocoding coordinates was larger than 50 km. In three of the four cases, this was caused by the presence of multiple cities with the same name in a country. In the fourth case, this was caused by an error in the WoS database. A small number of WoS publications have an address in Riyadh in South Africa. This should be Riyadh in Saudi Arabia. The four addresses with incorrect geocoding coordinates are all relatively unimportant. Together, the addresses occur in 343 publications.

**2.2. Indicators**

---

[3] The distinction between cities and provinces is not always clearly indicated in an address. What we refer to as cities may therefore sometimes be provinces. In the case of US and Canadian addresses, we took into account not only the city and the country indicated in an address but also the state or the province. State or province information seems to be provided consistently in a large majority of the US and Canadian publications.

Using the results of our geocoding procedure, we calculated the *geographical collaboration distance* (GCD) of each selected publication. We define the GCD of a publication as the largest geographical distance between two addresses mentioned in the publication's address list.[4] If a publication's address list contains only one address, the GCD of the publication is defined as zero. As mentioned earlier, publications that do not have any address at all were not taken into consideration in our analysis. Due to the limitations of the geocoding procedure, the coordinates of some of the addresses of a publication may be unknown. This turned out to be the case for 2.3% of the selected publications. For these publications, the addresses with unknown coordinates were disregarded and the GCD was calculated based on the remaining addresses. Notice that this may cause the GCD of these publications to be biased downwards.

Based on the GCD of a publication, we define the following four indicators of scientific globalisation:

- *Mean geographical collaboration distance* (MGCD): Average GCD of a set of publications;
- *Percentage medium- and long-distance collaborations* (%MLDC): Percentage of publications with a GCD of more than 200 km;
- *Percentage long-distance collaborations* (%LDC): Percentage of publications with a GCD of more than 1000 km;
- *Percentage very long-distance collaborations* (%VLDC): Percentage of publications with a GCD of more than 5000 km.

These indicators can be calculated for any set of publications as defined according to some delineation criterion, either geographical (e.g., country, region or city), institutional (e.g., university, research institute or company) or cognitive (e.g., field of science or research topic).

When counting publications and calculating our indicators, we take a fractional counting approach. For instance, a publication with addresses from two countries is treated as belonging half to each country. The use of a fractional counting approach ensures that statistics calculated at lower aggregations levels (e.g., country or field of science) can be directly compared with statistics calculated at higher aggregation levels (e.g., all countries or all fields of science).

## 3. Results

### 3.1. Overall statistics

Science has globalised at a fairly steady rate. The MGCD for science as a whole has increased more or less linearly over the past three decades from 334 km in 1980 to 1 553 km in 2009 (see Figure 1, left panel).[5] Between 2000 and 2009, the average growth per year was 47 km (corresponding with an average annual growth rate of 3.6%). It is important to realise that the WoS database is continuously expanding by adding new journals to its coverage, often journals with a local or regional focus. This could influence our results. However, checking our results with a fixed journals version of WoS, consisting only of journals that were permanently indexed between 2000 and 2009, our results remain similar: An MGCD of 1 633 km in 2009 and an average growth of 52 km per year during the last decade (corresponding with an average annual growth rate of 3.8%).

---

[4] Alternatively, we could have defined the GCD of a publication as the average geographical distance between all pairs of addresses mentioned in the publication's address list. A drawback of this definition would have been that the GCD of a publication may depend heavily on various details of the way in which address data are processed. For instance, if a publication has two or more addresses in the same city (perhaps even belonging to the same organisation), are these addresses treated as one single address or as multiple different addresses? Because of issues such as these, we prefer to define the GCD of a publication as the largest geographical distance between two addresses mentioned in the publication's address list. We note that the main trends and patterns reported in this paper are not very sensitive to the exact way in which the GCD of a publication is defined.

[5] As can be seen in Figure 1, there is a sudden increase in the growth of the MGCD around 1990. We suspect this sudden increase to be a database artefact rather than a true effect. It may be that around 1990 there has been some change in the way in which addresses are recorded in the database.

The increase in collaboration distances occurred in different speeds depending on geographical scales (see Figure 1, right panel). The share of medium- and long-distance collaborations (%MLDC) has grown by more than a factor three between 1980 and 2009, and the share of long-distance collaborations (%LDC) has grown by almost a factor four. The fraction of very long-distance partnerships (%VLDC) has increased almost fivefold. Hence, collaboration occurs more and more across large distances.

The growth in collaboration distances between 1980 and 2009 correlates with various other developments. The share of co-publications (defined as publications with more than one address) within the WoS database has jumped from 27% to 62% during the last three decades (see Figure 2, left panel). The share of international co-publications has increased from 5% to 21%. The average number of authors per publication has risen from 2.5 to 4.5 (see Figure 2, right panel; see also Wuchty, Jones, & Uzzi, 2007).

**3.2. Statistics at the level of fields of science**

The evolution of world science in recent decades is not only driven by socio-economic and political factors, but also by the cognitive dynamics within scientific fields, such as the rise of the biomedical sciences, nanoscience and ICT. We distinguish between four broad fields of science: *Engineering Sciences and Technology* (ET), *Medical Sciences, Life Sciences and Agricultural Sciences* (MLA), *Natural Sciences, Computer Sciences and Mathematics* (NCM) and *Social Sciences, Humanities and Arts* (SHA). These fields were obtained by a grouping of WoS journal subject categories. Each subject category, comprising of a disciplinary coherent set of journals, was assigned to one of the four broad fields.

Figure 2 captures the differences among the four broad fields of science. NCM was and still is the most globalised of these four. This is partly the result of a long tradition of cross-border, resource-intensive 'big science' collaboration (especially in high-energy physics and astronomy), in which large research facilities and joint resources are shared by scientists spread across the globe. MLA, however, with only two-third of NCM's MGCD in 1980, has almost caught up with NCM's level of globalisation in 2009. ET was engaged in the same catching-up process, but has not been able to keep up with MLA's steep growth rate since 2003. In contrast, recent years have shown a remarkable increase within SHA, the field least prone to 'big science' teamwork and collaboration between individual researchers and scholars. SHA is still significantly behind the others but is closing in fast on ET. Note that ET and NCM show signs of declined growth rates in long-distance collaboration in the last decade, the reasons for which warrant further research, but may perhaps reflect a negative effect of 9/11 on international research programmes and intercontinental travel.

MGCD statistics for 35 smaller fields are reported in Table 1. The fields are listed in decreasing order of their MGCD in 2009. The most globalised fields are *Astronomy* *and Astrophysics* (MGCD of 4 301 km in 2009) and *Earth Sciences* (2 527 km), which are both acknowledged 'big science' domains. As expected, the bottom of the list is occupied by fields within the humanities: *Creative Arts, Culture and Music* (301 km) and *Literature* (109 km). We find surprisingly large MGCDs for *Statistical Sciences* (1 978 km) and *Economics and Business* (1 939 km) considering the apparent lack of extensive international research programmes or large joint facilities in these fields.

### 3.3. Statistics at the level of countries

Location matters in science. The distances between research partners are obviously influenced by the geographical location of research sites. Researchers and scholars based in geographically peripheral countries, regions or continents are more inclined to engage in long-distance partnerships. The effect of a country's location on the globe is aptly illustrated in Figure 4, which displays 113 colour-coded countries according to their MGCD in the period 2007–2009. Each of these countries has an output of at least 200 WoS publications in this period. As expected, peripheral countries in the southern hemisphere are characterised by the largest collaboration distances. New Zealand is an extreme case with an MGCD of 4 069 km. Several developing countries in the tropics also surpass the 4 000 km mark, owing to long-distance partners in either the northern or the southern hemisphere.

Detailed statistics for selected countries are reported in Tables 2 to 5. This selection is limited to research-intensive countries with an output of at least 3 000 WoS publications in 2009. Several of the world's leading science nations, such as the United Kingdom, the US and Germany, are also among the fastest globalising countries (see Table 5). In contrast, some of the 'catching up' countries, such as Iran, China, Turkey and Brazil, which are experiencing a rapid growth in publication output, have hardly any increase or even a decrease in their MGCD (see Table 4).[6] Apparently, the increase in long-distance collaboration and global networking does not keep pace with the rapid expansion, often from a low base-line, of their science systems and associated publication output (see also Royal Society, 2011, Section 2.1.1; Zhou & Glänzel, 2010). This effect, reflecting the focus of developing countries on building local research capabilities, is also found when the MGCD is calculated within a fixed journals version of WoS in which serials newly added to the database are not considered.

The current situation among African nations offers an interesting case in point on how structural country-level factors may affect patterns of international scientific collaboration and the globalisation rate of national science systems (cf. Royal Society, 2011, Section 2.2). Looking at the most actively publishing countries on the African continent, two major groups can be distinguished:
- Algeria, Morocco and Tunisia. These are traditional Francophone countries, oriented towards France and other Mediterranean European countries. They have a relatively low MGCD. Their MGCD growth between 2000 and 2009 is low or negative, and their share of international co-publications does not increase.
- Cameroon, Ghana, Kenya, South Africa, Tanzania and Uganda. Each of these

---

[6] This may also be part of the explanation of the declining MGCD growth observed for some scientific fields in Figure 2. Countries with a rapidly growing publication output tend to have a more or less stable or even a decreasing MGCD. Because the share of these countries (especially China) in the worldwide publication output grows over time, their non-increasing MGCD has a negative effect on the worldwide MGCD growth.

Sub-Saharan countries has a high MGCD as well as a high MGCD growth. Most of them have an Anglophone background. These countries house international research institutes and partner with a large variety of English language countries, resulting in the production of relatively large quantities of international co-publications.

Clearly, an Anglophone colonial history and concomitant opportunities for easier access to English speaking countries has a significant effect on the globalisation potential of a country. However, other socio-economic factors may also exert significant impacts on this potential, as indicated by Nigeria. This West African country with an Anglophone history has a relatively low MGCD as well as a stagnating growth in terms of MGCD and share of international co-publications.

### 3.4. Effect of the geographic dispersion of scientific research

Increases in MGCD may arise from two spatial processes. On the one hand, MGCD may increase because researchers are more willing or more able to collaborate, especially across longer distances. On the other hand, MGCD may rise simply because the locations where research is being done are becoming more dispersed across the globe. For instance, the increasing scientific activity in countries such as China and Brazil introduces a wider distribution of research sites. MGCD growth may be a natural consequence of such a geographically more dispersed scientific world, independent of researchers' propensity to engage in (long-distance) collaboration.

To analyse the effect of geographical dispersion of scientific activities, we measure the average distance between two randomly selected addresses in the WoS database. Hence, our indicator of geographical dispersion is calculated as

$$\text{dispersion} = \frac{\sum_i \sum_j n(i)n(j)d(i,j)}{\sum_i \sum_j n(i)n(j)},$$

where $n(i)$ denotes the number of times address $i$ occurs in the database and $d(i, j)$ denotes the distance between addresses $i$ and $j$. This dispersion measure has a low value if scientific research is concentrated in a small area and a high value if scientific research takes place all over the earth's surface.

Our findings, summarised in Table 6, indicate a mere 0.5% annual increase in geographical dispersion over the years 1980–2009, which is minor compared with the 5.4% annual growth in MGCD. Hence, although scientific activity has become more geographically dispersed over the last 30 years, this seems to explain only a small part of the increasing MGCD. The largest part of the increasing MGCD must be due to researchers becoming more willing or more able to participate in long distance collaborations. During the last decade, however, there has been a subtle shift, marked by an increased dispersion growth of 0.7% annually and a decreased MGCD growth of 3.6%. This shift may signal structural changes within the geographical architecture of world science, notably the emergence of new locations for partnering, such as the BRICS countries (Brazil, Russia, India, China and South Africa).

## 4. Conclusion

Geocoding the millions of affiliate addresses on publications in the scientific literature has opened up a rich source of empirical data on collaboration patterns and trends in science. Our collaboration distance measures can be calculated for any aggregate of publications. This enables a wide variety of measurements spanning the entire geographical scale from countries, as presented in this paper, to intra-national regions, urban agglomerates and cities. Similarly, applying disciplinary classification schemes allows for analyses at different cognitive levels of detail, ranging from broad fields of science down to small research areas, individual scientific journals and other tailored sets of publications.

Our focus in this paper has been on the macro-level structure and dynamics of the worldwide science system. We have found that in thirty years time there has been an almost fivefold increase in the average collaboration distance per publication. This increase has taken place throughout the whole of science. During the last decade, however, the growth in collaboration distances has declined somewhat in the natural sciences and the engineering sciences. Especially the social sciences have shown a fast growth during this period. There also turn out to be substantial differences among countries. Collaboration distances have increased fastest for traditional science nations. Catching up countries, such as China and Brazil, have a rapidly increasing publication output, but the growth in collaboration distances is small or even negative for these countries.

Our findings show an evolution from a system of loosely connected 20th-century nation-state science bases into a 21st-century interconnected and internationally networked global science system, characterised by increasingly large distances between research partners. It is anyone's guess when this development of scientific globalisation will reach its limit in terms of a stable global scientific collaboration structure.

## Acknowledgements

We thank Cornelis van Bochove (CWTS) for critical discussion of intermediate results, Will Felps (Rotterdam School of Management, Erasmus University) for pointing us to some relevant literature and Suze van der Luijt (CWTS) for assistance in data pre-processing.## References

Florida, R. (2005). The world is spiky. *The Atlantic Monthly*, 48–51.
Frenken, K., Hardeman, S., & Hoekman, J. (2009). Spatial scientometrics: Towards a cumulative research program. *Journal of Informetrics*, *3*(3), 22–232.
Glänzel, W. (2001). National characteristics in international scientific co-authorship relations. *Scientometrics*, *51*(1), 69–115.
Hoekman, J., Frenken, K., & Tijssen R.J.W. (2010). Research collaboration at a distance: Changing spatial patterns of scientific collaboration within Europe. *Research Policy*, *39*(5), 662–673.
Hoekman, J., Frenken, K., & Van Oort, F. (2009). The geography of collaborative knowledge production in Europe. *Annals of Regional Science*, *43*(3), 721–738.
Katz, J.S. (1994). Geographical proximity and scientific collaboration. *Scientometrics*, *31*(1), 31–43.
Katz, J.S., & Martin, B.R. (1997). What is research collaboration? *Research Policy*, *26*(1), 1–18.
Leydesdorff, L., & Persson, O. (2010). Mapping the geography of science: Distribution patterns and networks of relations among cities and institutes. *Journal of the American Society for Information Science and Technology*, *61*(8), 1622–1634.
Liang, L., & Zhu, L. (2002). Major factors affecting China's inter-regional research8

Table 1. MGCD statistics per field of science.

| Field | MGCD 2009 | Annual growth MGCD 2000–2009 | Annual growth rate MGCD 2000–2009 |
|---|---|---|---|
| Astronomy and Astrophysics | 4 301 | 110 | 3.0% |
| Earth Sciences and Technology | 2 527 | 76 | 3.6% |
| Multidisciplinary | 2 371 | 71 | 3.6% |
| Statistical Sciences | 1 978 | 37 | 2.1% |
| Economics and Business | 1 939 | 60 | 3.7% |
| Environmental Sciences and Technology | 1 800 | 77 | 5.6% |
| Basic Life Sciences | 1 799 | 48 | 3.1% |
| Biological Sciences | 1 742 | 56 | 3.9% |
| Computer Sciences | 1 708 | 52 | 3.6% |
| Mathematics | 1 707 | 15 | 0.9% |
| Management and Planning | 1 635 | 60 | 4.5% |
| Physics and Materials Science | 1 619 | 19 | 1.2% |
| Biomedical Sciences | 1 595 | 54 | 4.1% |
| Basic Medical Sciences | 1 510 | 55 | 4.5% |
| Electrical Engineering and Telecommunication | 1 487 | 45 | 3.6% |
| Psychology | 1 478 | 64 | 5.7% |
| Civil Engineering and Construction | 1 428 | 40 | 3.3% |
| Clinical Medicine | 1 428 | 59 | 5.3% |
| Health Sciences | 1 421 | 71 | 6.9% |
| Agriculture and Food Science | 1 395 | 52 | 4.7% |
| Instruments and Instrumentation | 1 331 | 33 | 2.9% |
| Social and Behavioral Sciences, Interdisciplinary | 1 318 | 70 | 7.5% |
| General and Industrial Engineering | 1 296 | 50 | 4.9% |
| Mechanical Engineering and Aerospace | 1 216 | 36 | 3.5% |
| Energy Science and Technology | 1 196 | 15 | 1.4% |
| Chemistry and Chemical Engineering | 1 141 | 33 | 3.4% |
| Information and Communication Sciences | 1 064 | 57 | 7.5% |
| Sociology and Anthropology | 1 063 | 55 | 7.3% |
| Educational Sciences | 969 | 43 | 5.8% |
| Political Science and Public Administration | 905 | 48 | 7.6% |
| Law and Criminology | 724 | 31 | 5.5% |
| Language and Linguistics | 710 | 19 | 3.2% |
| History, Philosophy and Religion | 401 | 26 | 10.5% |
| Creative Arts, Culture and Music | 301 | 25 | 16.0% |
| Literature | 109 | 9 | 17.4% |
| All fields | 1 553 | 47 | 3.6% |

Table 2. Publication output and MGCD statistics for the top 10 countries with the largest output.

| Country | 2009 | | Annual growth rate 2000–2009 | |
|---|---|---|---|---|
| | Output | MGCD | Output | MGCD |
| US | 271 383 | 1 883 | 1.5% | 4.7% |
| China | 108 202 | 1 302 | 17.1% | 1.1% |
| Japan | 64 362 | 1 152 | -0.4% | 3.4% |
| United Kingdom | 63 355 | 1 681 | 0.5% | 6.7% |
| Germany | 61 290 | 1 360 | 1.3% | 4.6% |
| France | 43 894 | 1 452 | 1.3% | 4.9% |
| Canada | 38 959 | 1 953 | 3.7% | 5.0% |
| Italy | 36 744 | 1 162 | 4.1% | 3.1% |
| India | 33 729 | 948 | 9.0% | 1.9% |
| Spain | 31 970 | 1 183 | 6.3% | 3.5% |
| All countries | 1 134 979 | 1 553 | 4.1% | 3.6% |

Table 3. Publication output and MGCD statistics for the top 10 countries with the largest MGCD.

| Country | 2009 | | Annual growth rate 2000–2009 | |
| --- | --- | --- | --- | --- |
| | Output | MGCD | Output | MGCD |
| New Zealand | 4 515 | 4 154 | 2.9% | 5.1% |
| Australia | 27 298 | 3 604 | 4.7% | 4.4% |
| Chile | 3 180 | 3 128 | 9.6% | 1.0% |
| South Africa | 5 098 | 2 898 | 6.2% | 3.9% |
| Singapore | 5 832 | 2 828 | 7.2% | 6.9% |
| Thailand | 3 450 | 2 674 | 16.4% | -1.2% |
| Argentina | 5 070 | 2 411 | 4.0% | 2.9% |
| Canada | 38 959 | 1 953 | 3.7% | 5.0% |
| Israel | 8 579 | 1 915 | 1.0% | 2.1% |
| US | 271 383 | 1 883 | 1.5% | 4.7% |
| All countries | 1 134 979 | 1 553 | 4.1% | 3.6% |

Table 4. Publication output and MGCD statistics for the top 10 countries with the largest output growth.

| Country | 2009 Output | 2009 MGCD | Annual growth rate 2000–2009 Output | Annual growth rate 2000–2009 MGCD |
|---|---|---|---|---|
| Iran | 12 547 | 806 | 30.4% | -2.6% |
| Malaysia | 3 344 | 1 541 | 19.9% | -2.2% |
| China | 108 202 | 1 302 | 17.1% | 1.1% |
| Turkey | 19 340 | 542 | 16.8% | -1.8% |
| Thailand | 3 450 | 2 674 | 16.4% | -1.2% |
| Romania | 4 930 | 642 | 15.1% | -4.9% |
| Brazil | 25 714 | 1 406 | 12.5% | -2.7% |
| South Korea | 31 673 | 1 112 | 11.4% | 1.7% |
| Portugal | 5 931 | 1 339 | 10.9% | 2.4% |
| Taiwan | 20 560 | 956 | 9.8% | 0.8% |
| All countries | 1 134 979 | 1 553 | 4.1% | 3.6% |

Table 5. Publication output and MGCD statistics for the top 10 countries with the largest MGCD growth.

| Country | 2009 | | Annual growth rate 2000–2009 | |
|---|---|---|---|---|
| | Output | MGCD | Output | MGCD |
| Ireland | 3 969 | 1 459 | 7.4% | 7.2% |
| Singapore | 5 832 | 2 828 | 7.2% | 6.9% |
| United Kingdom | 63 355 | 1 681 | 0.5% | 6.7% |
| Norway | 5 876 | 1 522 | 5.3% | 5.5% |
| New Zealand | 4 515 | 4 154 | 2.9% | 5.1% |
| Canada | 38 959 | 1 953 | 3.7% | 5.0% |
| France | 43 894 | 1 452 | 1.3% | 4.9% |
| Switzerland | 12 788 | 1 765 | 2.6% | 4.8% |
| US | 271 383 | 1 883 | 1.5% | 4.7% |
| Germany | 61 290 | 1 360 | 1.3% | 4.6% |
| All countries | 1 134 979 | 1 553 | 4.1% | 3.6% |

Table 6. Geographical dispersion versus MGCD.

|  | Geographical dispersion | MGCD |
|---|---:|---:|
| 1980 | 6 031 km | 334 km |
| 2000 | 6 554 km | 1 131 km |
| 2009 | 7 008 km | 1 553 km |
| Annual growth rate 1980–2009 | 0.5% | 5.4% |
| Annual growth rate 2000–2009 | 0.7% | 3.6% |

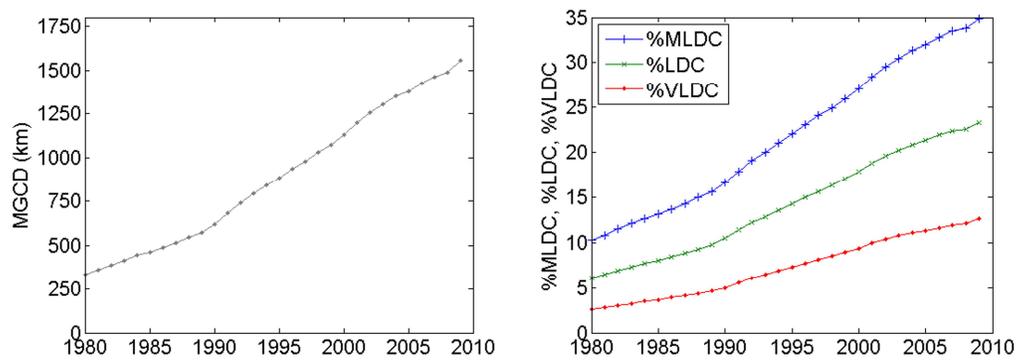

Figure 1. Trend in collaboration distance for science as a whole: MGCD (left panel) and %MLDC, %LDC and %VLDC (right panel).

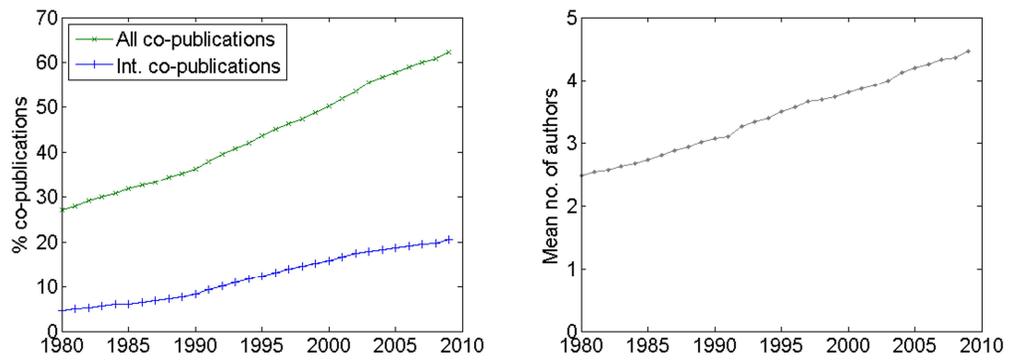

Figure 2. Trend in percentage co-publications (left panel) and average number of authors per publication (right panel).

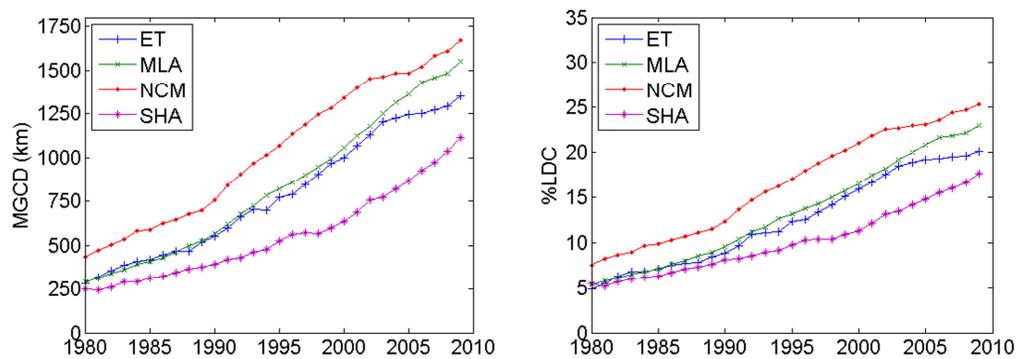

Figure 3. Trend in collaboration distance for four broad fields of science: MGCD (left panel) and %LDC (right panel).

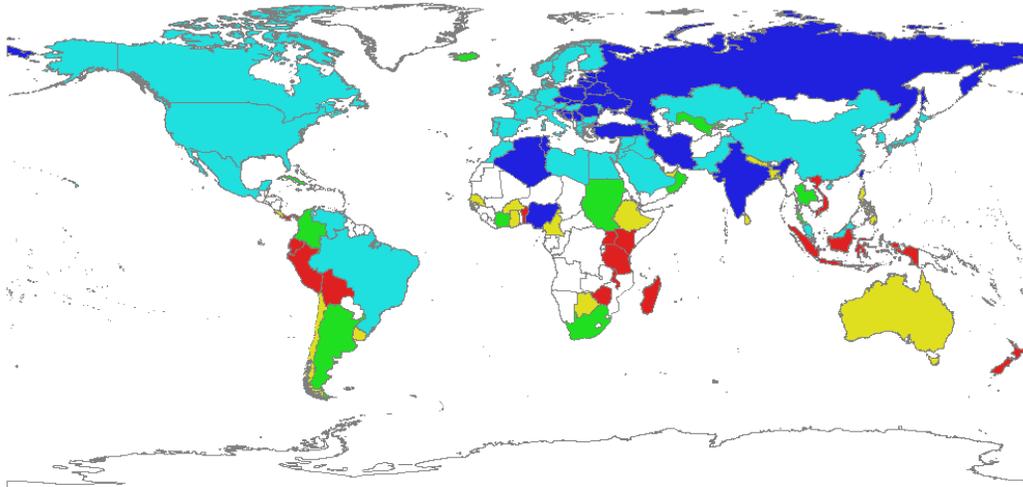

Figure 4. World map with colours indicating countries' MGCD in the period 2007–2009. Colour coding: Dark blue: MGCD < 1 000 km; Light blue: MGCD between 1 000 and 2 000 km; Green: MGCD between 2 000 and 3 000 km; Yellow: MGCD between 3 000 and 4 000 km; Red: MGCD > 4 000 km; White: Fewer than 200 publications, no MGCD calculated.